\documentclass[aps,pra, twocolumn, showpacs]{revtex4}

\usepackage{graphicx}
\usepackage{amsmath}
\usepackage{bm}

\graphicspath{{./IMAGES/}}

\begin{document}
\title{Two-electron atom with a screened interaction}

\author{C. A. Downing}
\email[]{downing@ipcms.unistra.fr}
\affiliation{Universit\'{e} de Strasbourg, CNRS, Institut de Physique et Chimie des Mat\'{e}riaux de Strasbourg UMR 7504, F-67034 Strasbourg, France}

\date{\today}

\begin{abstract}
We present analytical solutions to a quantum mechanical three-body problem in three dimensions, which describes a helium-like two-electron atom. Similarly to Hooke's atom, the Coulombic electron-nucleus interaction potentials are replaced by harmonic potentials. The electron-electron interaction potential is taken to be both screened (decaying faster than the inverse of the inter-particle separation) and regularized (in the limit of zero separation). We reveal the exactly solvable few-electron ground-state, which explicitly includes electron correlation, for certain values of the harmonic containment.
\end{abstract}

\maketitle

\section{\label{intro}Introduction}

Few-particle systems in quantum mechanics, from two-electron atoms \cite{Bethe2008, Maruhn2010} and simple molecules \cite{Pauling1935}, to electrons in a magnetic field \cite{Laughlin1983} and anyon excitons \cite{Portnoi1996, Parfitt2003}, are of great importance due to their ability to shed light on the many-body problem without recourse to numerics.

The helium atom \cite{Tanner2000} can be nicely treated semi-classically \cite{Ezra1991}, but the strong motivation to discover exact, quantum mechanical solutions led to research into artificial two-electron atoms. The prototypical artificial atomic model is Hooke's atom \cite{Kestner1968, Kais1989, Kais1993, Taut1993}, sometimes refereed to as hookium or harmonium. In this scenario, the Coulombic electron-nucleus interaction potential is replaced by a harmonic confining field. Remarkably, such a prescription is justified for two electrons in a quantum dot \cite{Kumar1990, Merkt1991}.

Unfortunately, analytical solutions only exist for a small number of other artificial helium-like atoms. Upon replacing the pure Coulombic electron-electron interaction, one may study: Moshinsky's atom \cite{Moshinsky1968, Dahl2009, Yanez2010, Bouvrie2012} which has a harmonic electron-electron interaction, Crandall's atom \cite{Crandall1984, Crandall1985, Makarewicz1988} with an inverse-square electron-electron interaction, or Samanta's atom \cite{Samanta1990, Ghosh1991, Karwoswki2014} with a combined linear and Coulombic interaction. We should also mention that recently more exotic geometries have been studied, including the so-called ballium atom \cite{Loos2010} or spherium atom \cite{Ezra1982, Loos2009, Loos2010c} where the electrons are constrained to move on the surface of a sphere, and where both singlet and triplet states have been discussed in detail \cite{Loos2010b}. 

These aforementioned beautiful atoms have been proven invaluable for detailed studies on few-electron entanglement \cite{Pipek2009, Benavides2012, Lin2013}, statistical and kinetic energy correlation effects \cite{Laguna2011, Amovilli2003} and the Fisher-Shannon information \cite{Nagy2006}, as well as rotations and stretching vibrations of molecules \cite{Yannouleas2000}.

In this work we introduce a quantum three-body problem which yields transparent, closed-form solutions for both singlet and triplet states. The two-electron atom we consider is noticeable for having a regularized electron-electron interaction, which also decays faster at large inter-particle separations then the unscreened Coulomb potential. Therefore our model does not display the Kato cusp phenomenon, which is an artefact of purely Coulombic interactions. We identify the simple form of the term in the exactly-solvable ground state wavefunction which accounts for electron correlation, and provide an analytical expression for the electron density. In our derivation of the two-electron wavefunction, we utilize the increasingly popular special functions of the confluent Heun class \cite{Ronveaux, Fiziev2009, Downing2013, Downing2016}.

\begin{figure}[tb] 
\includegraphics[width=0.45\textwidth]{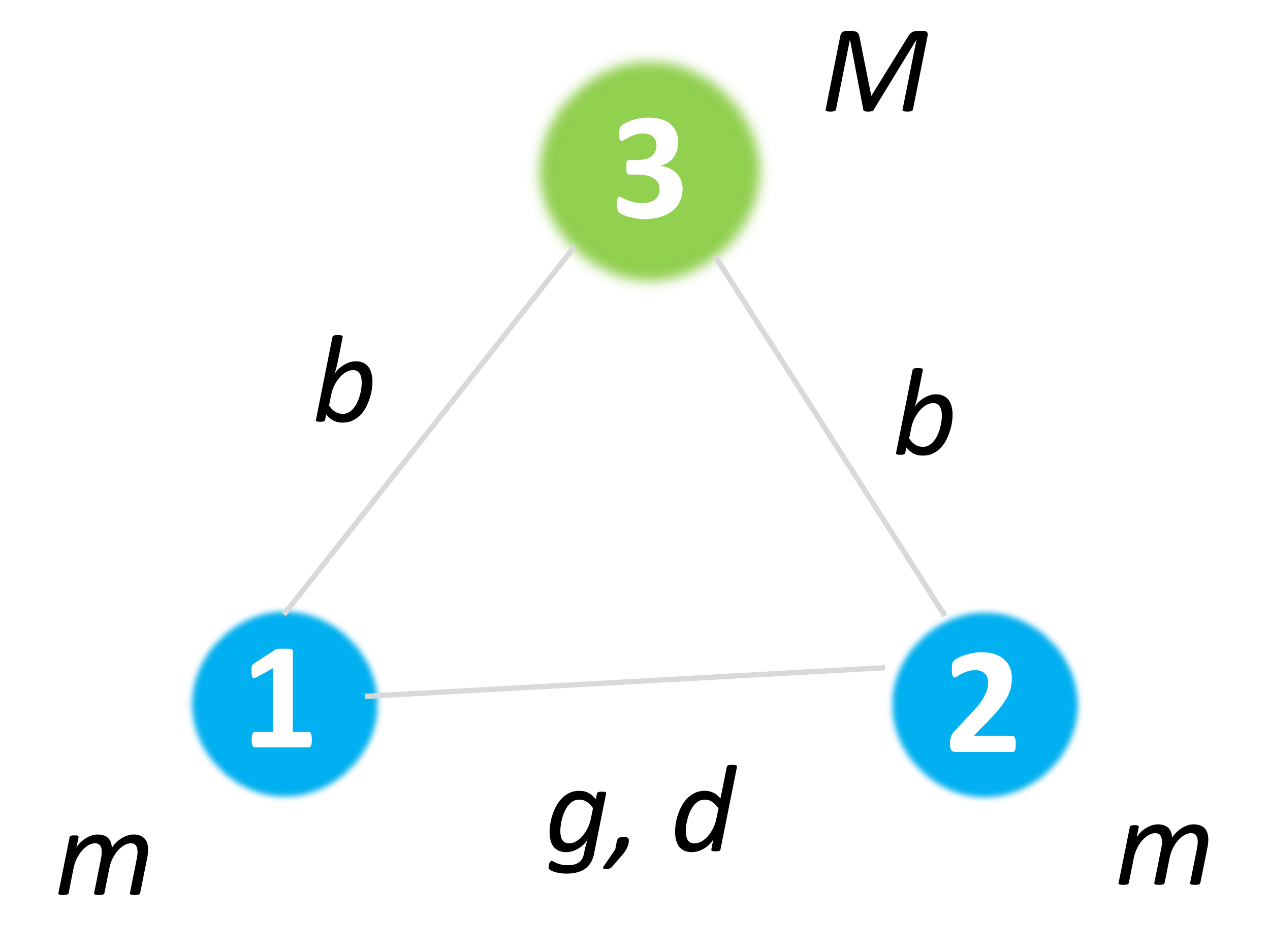}  
\caption{(Color online) A sketch of the three-body problem under consideration, comprised of particle $1$ of mass $M$ and particles $2$ and $3$, both of mass $m$. The harmonic interactions $V_1$ and $V_2$ are associated with the parameter $b$, whilst the electron-electron interaction $V_3$ is defined by two parameters, $g$ and $d$.}
\label{fig1}
\end{figure}

We consider a system of three interacting particles, as depicted in Fig.~\ref{fig1}, governed by the following Hamiltonian (written in atomic units $\hbar = m = e = 1$)
\begin{multline}
\label{intro1}
 H = -\frac{1}{2} \left( \nabla_1^2 + \nabla_2^2 \right) - \frac{1}{2 M} \nabla_3^2 \\ + V_1 \left( |\boldsymbol{r}_1 - \boldsymbol{r}_3| \right) + V_2 \left( |\boldsymbol{r}_2 - \boldsymbol{r}_3| \right) + V_3 \left( |\boldsymbol{r}_1 - \boldsymbol{r}_2| \right),
\end{multline}
where particles $1$ and $2$ are both of mass $m$, representing electrons, and particle $3$ is of mass $M$, representing the nucleus, such that the system is a helium-like two-electron atom. The rest of this work is devoted to solutions of Eq.~\eqref{intro1} and is organized as follows. We present the specific Hamiltonians for our chosen interactions in Sec.~\ref{three}, where we also provide the solutions to the more straightforward equations separated from the full Schr\"{o}dinger equation. In Sec.~\ref{qes}, we derive exact solutions to the remaining nontrivial equation arising from the relative motion of the electrons. We unveil the analytical form of the two-electron ground state and give a measure of the correlation via the electron density in Sec.~\ref{corr}. Finally, we draw some conclusions in Sec.~\ref{conc}.

\section{\label{three}The three-body problem}

The three body Hamiltonian~\eqref{intro1} can be most conveniently written down with the coordinate transformation $(\boldsymbol{r}_1, \boldsymbol{r}_2, \boldsymbol{r}_3) \to (\boldsymbol{R}, \boldsymbol{S}, \boldsymbol{r})$, where
\begin{subequations}
\label{three1}
 \begin{align}
  \boldsymbol{R} = \frac{1}{2 + M} \left( \boldsymbol{r}_1 + \boldsymbol{r}_2 + M \boldsymbol{r}_3  \right),  \label{co1} \\
  \boldsymbol{S} = \frac{1}{\sqrt{2}} \left( \boldsymbol{r}_1 + \boldsymbol{r}_2 - 2 \boldsymbol{r}_3 \right), \label{co2} \\
  \boldsymbol{r} = \frac{1}{\sqrt{2}} \left( \boldsymbol{r}_1 - \boldsymbol{r}_2 \right). \label{co3}
 \end{align}
\end{subequations}
Then the Hamiltonian~\eqref{intro1} converts into the form
\begin{multline}
\label{three2}
 H =  - \frac{1}{2 M}  \frac{1}{1 + \tfrac{2}{M}}  \nabla_{\boldsymbol{R}}^2 -\frac{1}{2} \left( 1 +  \tfrac{2}{M} \right)  \nabla_{\boldsymbol{S}}^2 -\frac{1}{2} \nabla_{\boldsymbol{r}}^2 \\
 V_1 \left( \tfrac{1}{\sqrt{2}} |\boldsymbol{S} + \boldsymbol{r}| \right) + V_2 \left( \tfrac{1}{\sqrt{2}} |\boldsymbol{S} - \boldsymbol{r}| \right) + V_3 \left( \sqrt{2} |\boldsymbol{r}| \right).
\end{multline}
We consider harmonic interactions between the particle $3$ and each of the two particles $1$ and $2$ respectively, a situation which may be realized in few electron quantum dots \cite{Coe2008}. Explicitly, with the inter-particle distances $r_{ij} = | \bm{r}_i - \bm{r}_j|$, we specify
\begin{equation}
\label{three3}
  V_1 \left( r_{13} \right) = \frac{r_{13}^2}{2 b^4},\quad V_2 \left( r_{23} \right) = \frac{r_{23}^2}{2 b^4}, 
\end{equation}
where we have introduced the parameter $b$ to govern the harmonic interaction strength. Such a choice of harmonic interactions enables the total Hamiltonian to be decomposed as $H = H_R + H_S + H_r$, describing the center-of-mass ($\boldsymbol{R}$), pseudorelative ($\boldsymbol{S}$) and relative ($\boldsymbol{r}$) motion respectively. We study the interaction between the particles $1$ and $2$ which is both screened (falling asymptotically faster than the Coulombic interaction) and regularized (in the limit of vanishing inter-particle separation). Namely, we choose
\begin{equation}
\label{three4}
 V_3 \left( r_{12} \right) =  \frac{g}{r_{12}^2 +  2 d^2},
\end{equation}
where $d$ modulates the spatial extent of the interaction and $g$ is a coupling constant. Notably, there are two limiting cases of interest within Eq.~\eqref{three4}: that of an inverse-square interaction potential $V_3 \propto 1/r_{12}^2$ when $d \to 0$, which recall's Crandall's atom \cite{Crandall1984}; and a harmonic interaction potential $V_3 \propto 1 - r_{12}^2/ 2 d^2$ when $d \to \infty$, which recovers Moshinsky's atom \cite{Moshinsky1968}.

Now, with these aforementioned choices of interaction~\eqref{three3}~and~\eqref{three4}, the three-particle system is governed by three distinct Hamiltonians
\begin{subequations}
\label{three5}
 \begin{align}
  H_R = - \frac{1}{2 M} \frac{1}{1 + \tfrac{2}{M}}  \nabla_{\boldsymbol{R}}^2, \label{co6} \\
  H_S = -\frac{1}{2} \left( 1 + \frac{2}{M} \right)  \nabla_{\boldsymbol{S}}^2 + \frac{1}{2 b^4} |\boldsymbol{S}|^2 , \label{co7} \\
  H_r = -\frac{1}{2} \nabla_{\boldsymbol{r}}^2  + \frac{1}{2 b^4} |\boldsymbol{r}|^2 + \frac{1}{2} \frac{g}{|\boldsymbol{r}|^2 + d^2}. \label{co8} 
 \end{align}
\end{subequations}
These Hamiltonians~\eqref{three5} are connected by separation constants which arise from the total energy $E$ in the full Schr\"{o}dinger equation, where $E = E_R + E_S + E_r$ describes the center-of-mass, pseudorelative and relative eigenenergies each in turn. Since the total Hamiltonian $H$ is independent of spin, the total wavefunction may be factorized as
\begin{equation}
\label{spin1}
 \Psi =  \Phi(\boldsymbol{R}) \varphi(\boldsymbol{S}) \psi(\boldsymbol{r}) \chi,
\end{equation}
where the symmetry of the spin factor $\chi$ accounts for both singlet ($\chi$ antisymmetric) and triplet ($\chi$ symmetric) wavefunctions, as is required by the Pauli exclusion principle. The Hamiltonians~\eqref{co6} and~\eqref{co7} are the familiar free particle and isotropic harmonic oscillator Hamiltonians respectively, with well-known eigenfunctions hereby denoted $\Phi(\boldsymbol{R})$ and $\varphi(\boldsymbol{S})$, and the following eigenvalues: 
\begin{subequations}
\label{three6}
 \begin{align}
  E_R = \frac{1}{2 M} \frac{|\boldsymbol{K}|^2}{1 + \tfrac{2}{M}} , \label{co9} \\
  E_S^{n_s, l_s} = \frac{1}{2 b^2} \left( 1 + \tfrac{2}{M} \right)^{1/2} \left( 3 + 4 n_s + 2 l_s \right). \label{co10} 
 \end{align}
\end{subequations}
Here $\boldsymbol{K}$ is the center-of-mass wavevector, and $n_s$ and $l_s$ are nonnegative integers arising due to the pseudorelative $(\boldsymbol{S})$ motion. The third quantum number $m_s$ does not enter the energy quantization Eq.~\eqref{co10}, as is known from the solution of the isotropic harmonic oscillator.

In the next section, we consider in detail the final Schr\"{o}dinger equation of the system formed from the Hamiltonian~\eqref{co8} and associated with the eigenvalue $E_r$. The relative $(\boldsymbol{r})$ motion also gives rise to three quantum numbers, designated $\left(n_r, m_r, l_r\right)$, such that the final eigensolution may be described by a total of six quantum numbers.

\section{\label{qes}Relative motion}

We solve the remaining Schr\"{o}dinger equation $H_r \psi = E_r \psi$ in spherical coordinates $(r, \theta, \phi)$. We choose the typical decomposition of $\psi$ utilizing spherical harmonics and some radial function to be determined, namely $\psi (\boldsymbol{r}) = Y^{l_r, m_r}(\theta, \phi) R(r)$. This ensures that the parity of the relative wavefunction satisfies Pauli's principle. The eigenfunctions $\psi$ belong to either singlet ($l_r$ even, hence $\psi$ symmetric) or triplet ($l_r$ odd, hence $\psi$ antisymmetric) total wavefunctions $\Psi$, see Eq.~\eqref{spin1}. The radial differential equation following from the Hamiltonian~\eqref{co8} is
\begin{equation}
\label{qes1}
 R'' + \frac{2}{r} R' + \left( k^2 - \frac{r^2}{b^4} - \frac{g}{r^2 + d^2} - \frac{l_r (l_r + 1)}{r^2} \right) R = 0,
\end{equation}
where $l_r$ is the azimuthal quantum number for relative motion, $k^2 = 2 E_r$ and $'$ denotes taking a derivative with respect to the independent variable. Since the radial function should behave like $R \sim r^{l_r}$ at short range $r \sim 0$, and due to the characteristic decay $R \sim \exp (-r^2/2 b^2)$ as $r \to \infty$, we undertake a peeling-off procedure of the asymptotics with the help of the ansatz
\begin{equation}
\label{qes4}
 R(z) = z^{\tfrac{l_r}{2}} e^{- \tfrac{z d^2}{2 b^2}} y(z),
\end{equation}
where we work in the dimensionless variable $z = (r/d)^2$ and $y(z)$ is an auxiliary function to be determined. We obtain from Eq.~\eqref{qes1} an equation for $y(z)$:
\begin{multline}
\label{qes5}
 z y'' + \left( l_r + \tfrac{3}{2} - z \tfrac{d^2}{b^2} \right) y' \\
 + \left( \tfrac{k^2 d^2}{4} - \tfrac{g/4}{1+z} - \tfrac{d^2}{2 b^2} \left( l_r + \tfrac{3}{2} \right) \right) y = 0.
\end{multline}
After the elimination of the independent variable with $z = -\xi$, and switching the dependent variable with $y(\xi) = (1-\xi) f(\xi)$, Eq.~\eqref{qes5} becomes precisely in the general form of a confluent Heun equation 
\begin{equation}
\label{qes7}
 f'' + \left( \alpha + \frac{\beta +1}{\xi} + \frac{\gamma +1}{\xi-1} \right) f' + \left( \frac{ \mu }{\xi} + \frac{\nu}{\xi-1} \right) f = 0.
\end{equation}
In the case of our treated atom, the parameters $(\alpha, \beta, \gamma, \delta, \eta)$ appearing in Eq.~\eqref{qes7} are found to be
\begin{subequations}
\label{qes8}
 \begin{align}
  \alpha = \tfrac{d^2}{b^2},~~~\beta = \tfrac{1}{2} + l_r,~~~\gamma = 1, \label{vit8}
 \\
    \mu = \tfrac{1}{4} \left( g - k^2 d^2 \right) + \left( l_r + \tfrac{3}{2} \right) \left( \tfrac{d^2}{2b^2} - 1 \right), \label{vit88}
\\
\nu = \tfrac{3}{2} + l_r + \tfrac{d^2}{b^2} - \tfrac{g}{4}. \label{vit888}
 \end{align}
\end{subequations}
With the popular parameter parameterization $(\mu, \nu) \to (\delta, \eta)$ \cite{Note1}, we obtain the useful parameters
\begin{equation}
\label{qes559}
  \delta = - \tfrac{1}{4} k^2 d^2,~~~\eta = \tfrac{1}{2} + \tfrac{1}{4} (k^2 d^2 - g).
\end{equation}
The confluent Heun equation~\eqref{qes7} has as its the solution the confluent Heun function \cite{Ronveaux}
\begin{equation}
\label{qes9}
 f(z) = H_C \left( \alpha, \beta, \gamma, \delta, \eta, \xi \right),
\end{equation}
which can be formally represented as a power series, with the radius of convergence $|\xi| < 1$, in the form:
\begin{equation}
\label{heunsol}
	H_C(\alpha, \beta, \gamma, \delta, \eta, \xi) = \sum\limits_{n=0}^\infty v_n(\alpha, \beta, \gamma, \delta, \eta, \xi) \xi^n.
\end{equation}
Here the coefficients $v_n$ are given by the three-term recurrence relation 
\begin{equation}
\label{relation}
 A_n v_n = B_n v_{n-1} + C_n v_{n-2}, \quad v_{-1} = 0, \quad v_{0} = 1,
\end{equation}
and the coefficients are given by
\begin{subequations}
\label{recurrence}
 \begin{align}
  A_n &= 1 + \frac{\beta}{n}, \label{conda} \\
  B_n &= 1 + \frac{\beta + \gamma - \alpha - 1}{n}  + 
				 \frac{\eta}{n^2} - \frac{\beta + \gamma - \alpha + \alpha \beta - \beta \gamma}{2 n^2} , \label{condb} \\
  C_n &= \frac{\alpha}{n^2} \left( \frac{\delta}{\alpha} + \frac{\beta + \gamma}{2} + n - 1 \right). 
 \end{align}
\end{subequations}
Therefore, using Eqs.~\eqref{qes4} and~\eqref{qes9}, one may write down the following solution to the radial equation~\eqref{qes1}
\begin{equation}
\label{qes13}
 R_N(z)  =  z^{l_r /2} (1+z) H_C \left( \alpha, \beta, \gamma, \delta, \eta, -z \right) e^{- \tfrac{z d^2}{2 b^2}}.
\end{equation}
Here $N =1, 2, 3,..$ is not the principal quantum number, but refers to an integer which arises upon terminating the confluent Heun function. Hence $N$ is displayed as a subscript to distinguish it from quantum numbers (which are displayed in superscripts throughout this work). This aforementioned termination requirement is necessary for Eq.~\eqref{qes13} to be square-integrable, hence the power series \eqref{heunsol} must reduce to polynomial. This is indeed the case for certain values of the parameters of the system $g$ and $d/b$. The two necessary conditions to terminate a confluent Heun function are
\begin{subequations}
\label{qes10}
 \begin{align}
   \frac{\delta}{\alpha} + \frac{\beta + \gamma}{2} + N + 1 = 0, \label{co49} \\
  \Delta_{N+1} = 0, \label{co449} 
 \end{align}
\end{subequations}
where $\Delta_{N+1}$ refers to a $(N+1)\times(N+1)$ tridiagonal matrix, the full form of which is written down in the Appendix of Ref.~\cite{Fiziev2009}. The first condition~\eqref{co49} is equivalent to the expression
\begin{equation}
\label{qes11}
 E_{r, N}^{l_r} = \frac{1}{2 b^2} \left( 7 + 2 l_r + 4N \right), 
\end{equation}
which defines the value of the eigenenergy for which one may write down the exact solution of the problem, starting with the threshold energy $E_{r, 1}^{0} = 11 / 2 b^2$. The second condition~\eqref{co49} may be written most compactly for the small matrices associated with $N=1, 2$ as follows
\begin{subequations}
\label{rtt}
 \begin{align}
   \Delta_{2} = 
\begin{vmatrix}
 \mu &  l_r + \tfrac{3}{2} \\
\tfrac{d^2}{b^2} &  \tfrac{d^2}{b^2} + \mu - l_r - \tfrac{7}{2}
\end{vmatrix}
	= 0, \label{firstcond} \\
  \Delta_{3} = 
\begin{vmatrix}
 \mu & l_r + \tfrac{3}{2} & 0 \\
 2 \tfrac{d^2}{b^2} & \tfrac{d^2}{b^2} + \mu - l_r - \tfrac{7}{2} & 2 l_r + 5\\
 0 & \tfrac{d^2}{b^2} & 2 \tfrac{d^2}{b^2} + \mu - 2 l_r - 9
\end{vmatrix}
	= 0, \label{secondcond}
 \end{align}
\end{subequations}
where we recall from Eq.~\eqref{vit88} that the second parameter $g$ enters via $\mu = \mu(g, d/b)$. The matrices $\Delta_{N+1}$ for higher values of $N$ may be constructed in a similar manner. Therefore, assuming both Eq.~\eqref{qes11} and Eq.~\eqref{rtt} are satisfied, one may write down the analytical solutions for the radial function (with $N=1, 2$) as 
\begin{subequations}
\label{3435}
 \begin{align}
  R_1^{n_r, l_r} &= \tfrac{\mathcal{N}_1^{n_r, l_r}}{d^{3/2}} \left( \tfrac{r}{d} \right)^{l_r} \left( 1+ \tfrac{r^2}{d^2} \right) \left( 1 - v_1 \tfrac{r^2}{d^2} \right) e^{- \tfrac{r^2}{2 b^2}}, \label{co456569} \\
  R_2^{n_r, l_r} &= \tfrac{\mathcal{N}_2^{n_r, l_r}}{d^{3/2}} \left( \tfrac{r}{d} \right)^{l_r} \left( 1+ \tfrac{r^2}{d^2} \right) \left( 1 - v_1 \tfrac{r^2}{d^2} + v_2 \tfrac{r^4}{d^4} \right) e^{- \tfrac{r^2}{2 b^2}}, \label{co4565690}
 \end{align}
\end{subequations}
in terms of the coefficients $v_n$ appearing in Eq.~\eqref{relation}. Here $n_r$ is the node number and $\mathcal{N}_N^{n_r, l_r}$ is a normalization constant. Taking into account the angular part of the wavefunction, $Y^{l_r, m_r}(\theta, \phi)$, the relative motion wavefunction $\psi$ is dependent on three quantum numbers $(n_r, m_r, l_r)$.

\begin{figure}[tb] 
\includegraphics[width=0.45\textwidth]{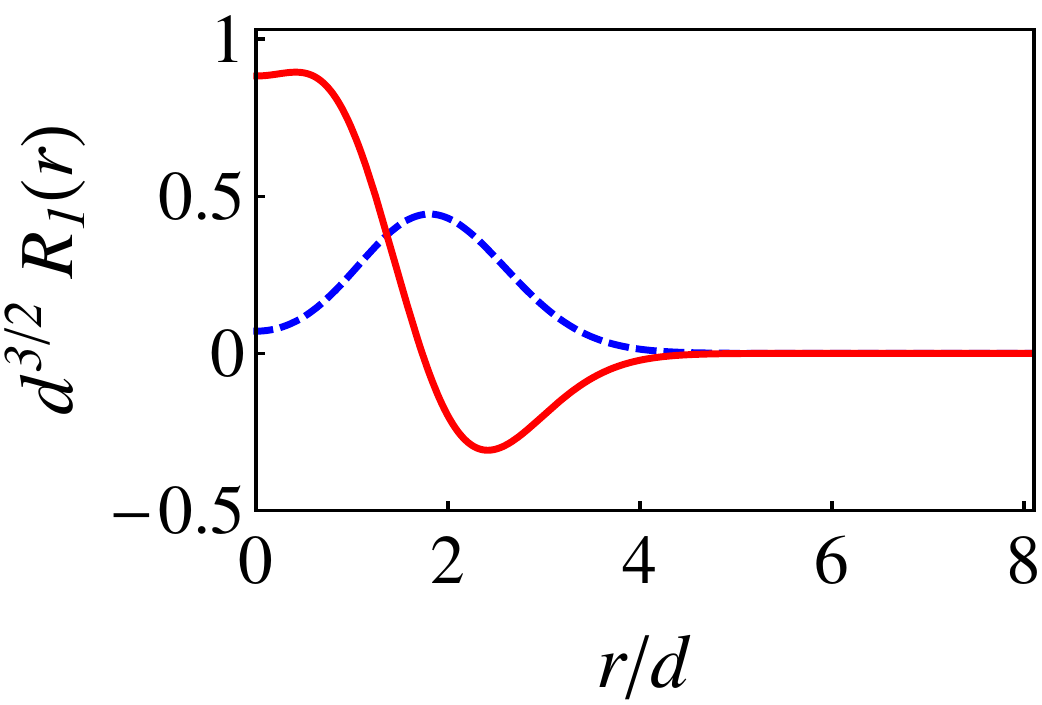}  
\caption{(Color online) A spatial plot of the $N=1, l_r =0$ radial function $R_1^{n_r, 0}$ given by Eq.~\eqref{co456569}, with parameter $d/b = 1$ and energy $E_{r, 1}^{0} = 11 / 2 b^2$. A ground state (with $g = 26, v_1=-2, n_r=0$) and an excited state (with $g = 12, v_1=1/3, n_r=1$) are shown as the dashed blue line and the solid red line respectively.}
\label{fig2}
\end{figure}

We plot in Fig.~\ref{fig2} the $N=1$ radial function~\eqref{co456569} for the simplest case of $l_r = 0$ (thus belonging to singlet states). It follows from Eq.~\eqref{qes11} that the associated eigenvalue is $E_{r, 1}^{0} = 11 / 2 b^2$. Upon choosing $d/b = 1$, one finds from the quadratic equation~\eqref{firstcond} two solutions $g = (26, 12)$, corresponding to the node-less ground state ($n_r = 0, g = 26$, dashed blue line) and a single-node excited state ($n_r = 1, g = 12$, solid red line) respectively. For completeness, the coefficients entering Eq.~\eqref{3435} in these circumstances are $v_1=(-2, 1/3)$ in turn. One notices in Fig.~\ref{fig2} both the anticipated nodal structure and the strong asymptotic decay of the radial functions due to the presence of dominant Gaussian functions.

The ground state radial function found $R_1^{0, 0}$ is most convenient for constructing the (singlet) ground state of the full wavefunction $\Psi$. Further radial functions $R_1^{n_r, l_r}$, suitable for building singlet ($l_r$ even) or triplet ($l_r$ odd) states, may be found in the same manner \cite{Note2}. Moreover, polynomial solutions of the class $N \ge 2$ may be assembled analogously to those of the type $N=1$ \cite{Note3}. An analysis of the solutions found with varying $N$ demonstrates the property that solutions of the class $N$ are $N+1$ in number and have a number of nodes $n_r$ ranging from a minimum of zero to a maximum of $N$. We should also mention the possibility to generate non-polynomial solutions for this problem, for example by following the method of constructing irrational solutions applied in similar systems \cite{Loos2012, Loos2012b}.

We shall now consider the two prescient limits of the system, which may be approached through Eq.~\eqref{three4}. In the regime of small spatial extent $d$, the interaction potential~\eqref{three4} is inverse-square in form. A direct mapping to the isotropic harmonic oscillator yields the eigenvalues
\begin{equation}
\label{qes2}
 E_r^{n_r, l_r} = \frac{1}{b^2} \left( 1 + 2 n_r + \left[ g + (l_r + \tfrac{1}{2})^2\right]^{\tfrac{1}{2}} \right),
\end{equation}
where $n_r$ is a nonnegative integer. Thus the exact ground state energy level in the limit of strong harmonic containment is $E_r^{0, 0} = (1 + \sqrt{g}) / b^2$. Degeneracies may sometimes occur at special values of the coupling constant $g$. If a degeneracy level does exist then
\begin{equation}
\label{deg}
 g =  \frac{\left[ \Delta l_r^2-4\Delta n_r^2 \right)]}{16 \Delta n_r^2} \left[ (l_r+l_r'+1)^2-4\Delta n_r^2 \right)],
\end{equation}
is satisfied, where $\Delta l_r = l_r-l_r'$ and $\Delta n_r = n_r-n_r'$. For example, the degenerate level $E_r^{1, 0} = E_r^{0, 3} = 5 / b^2$ appears with the coupling constant $g = 15/4$, whilst when $g = 10$ one finds the doubly degenerate level $E_r^{1, 1} = E_r^{0, 4} = 13 / 2 b^2$.

In the opposing limit of large $d$, the case of weak harmonic containment, the interaction potential~\eqref{three4} is of a (constant minus) quadratic form. Therefore, the harmonic oscillator strength is gently renormalized by the factor $\gamma = (1-g b^4/d^4) > 0$ and the eigenvalues are
\begin{equation}
\label{qes3}
 E_r^{n_r, l_r} = \frac{g}{2 d^2} + \frac{\gamma^{1/2}}{2 b^2} \left( 3 + 4 n_r + 2 l_r \right).
\end{equation}
Ignoring the constant energy shift, the ground state energy of the atom is $E_r^{0, 0} = 3 \gamma^{1/2}/ 2 m b^2$. Degenerate energy levels occur whenever 
\begin{equation}
\label{degg}
 2 (n_r - n_r') + (l_r - l_r') = 0,
\end{equation}
is satisfied. Therefore, each energy level $E_r^{n_r, l_r}$ is paired with levels like $E_r^{n_r', l_r + 2 (n_r - n_r')}$, which is valid as long as $n_r' \le n_r + l_r /2$. For example, one finds the doubly-degenerate level $E_r^{1, 0} = E_r^{0, 2} = 7 \gamma^{1/2}/ 2 b^2$ and the triply-degenerate level $E_r^{2, 0} = E_r^{1, 2} = E_r^{0, 4} = 11 \gamma^{1/2}/ 2 b^2$.

\section{\label{corr}The ground state}

We consider the limit of a two-particle system, such that $M \gg m$. In the center of mass of the system, the few-particle (singlet) ground state $(n_s = l_s = m_s = n_r = l_r = m_r = 0)$ has, from Eq.~\eqref{co10} and Eq.~\eqref{qes11}, the ground state energy $E^{\text{gs}} = 7 / b^2$. The associated $N=1$ ground state wavefunction is 
\begin{multline}
\label{corr1}
 \Psi_{1}^{\text{gs}} \left( \boldsymbol{r}_1, \boldsymbol{r}_2 \right) = \frac{1}{2 \pi^{5/4}} \frac{\mathcal{N}}{(b d)^{3/2}} \left( 1 + \frac{r_{12}^2}{2 d^2} \right) \\
 \times \left( 1 - \frac{v_1 r_{12}^2}{2 d^2} \right) e^{- \frac{1}{2 b^2} \left( r_1^2 + r_2^2 \right)},
\end{multline}
where $\mathcal{N}$ is a normalization constant. The electron correlation appears in the two bracketed factors containing the inter-electronic separation $r_{12}$. Most notably, the wavefunction~\eqref{corr1} does not possess a linear term in $r_{12}$ and hence does not exhibit a Kato cusp \cite{Kato1957}, which is fundamental to most atomic and molecular wavefunctions. This peculiarity is due to the absence of a purely Coulombic interaction between particles.

\begin{figure}[tb] 
\includegraphics[width=0.45\textwidth]{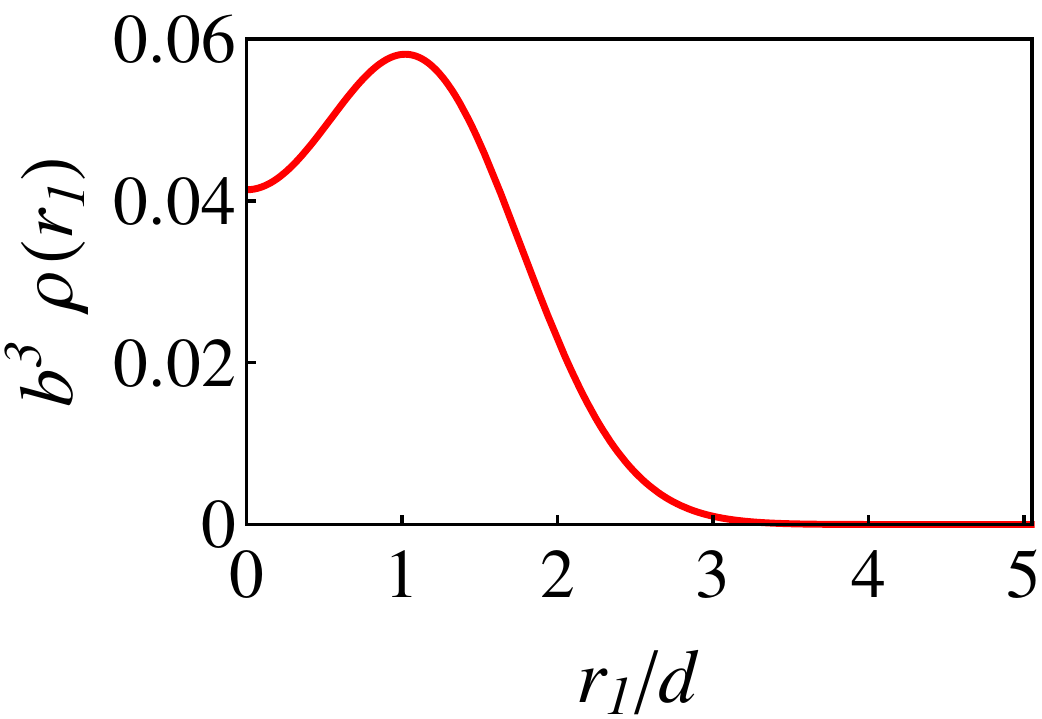}  
\caption{(Color online) A plot of the ground state electron density, given by Eq.~\eqref{corr2}, for the case of the $N=1$ two-electron wavefunction~\eqref{corr1}. The ground state energy is $E = 7 / b^2$, and the system parameters are $d/b = 1$, $g = 26$ and $v_1 = -2$.}
\label{fig3}
\end{figure}

The ground state electronic density $\rho (\boldsymbol{r}_1) = 2 \int |\Psi_{1} \left( \boldsymbol{r}_1, \boldsymbol{r}_2 \right)|^2 \mathrm{d} \boldsymbol{r}_2$ may be calculated analytically, yielding the result
\begin{align}
\begin{split}
\label{corr2}
 \rho (\boldsymbol{r}_1) = \frac{\mathcal{N}^2}{b^3} \frac{e^{- \frac{r_1^2}{b^2}}}{512 \pi} \left( \frac{b}{d} \right)^{11}  \left( 945 v_1^2 + 32 \left( \tfrac{d}{b} \right)^{6} f_1   \vphantom{\frac{d}{e}}\right. \\ 
 + 24 \left( \tfrac{d}{b} \right)^{4} f_2 + 16 \left( \tfrac{d}{b} \right)^{8} (2+y^2)^2 (v_1 y^2 -2)^2 \\ 
      \left.  + 840 v_1 \left( \tfrac{d}{b} \right)^{2} (3 v_1 y^2 + v_1 -1)  \vphantom{\frac{d}{e}}\right)
\end{split} 
\end{align}
with the auxiliary functions
\begin{multline*}
f_1 = 12 (v_1-1) + 10 [1 + v_1 (v_1 - 4)] y^2 + 21 v_1 (v_1-1) y^4 + 9 v_1^2 y^6, \\
f_2 = 10 + v_1 \left[ 10 (v_1-4) + 70 (v_1-1) y^2 + 63 v_1 y^4 \right],
\end{multline*}
where we have introduced the scaled variable $y = r_1/d$ and where the constraint $\int \rho (\boldsymbol{r}_1) \mathrm{d} \boldsymbol{r}_1 = 2$ fixes $\mathcal{N}$. We plot this spherically symmetric electronic density~\eqref{corr2} in Fig.~\ref{fig3}, which shows that the one-body density is a smooth function, starting with regular behavior at the origin and continuing up to a maximum at moderate range - a so-called fat attractor \cite{Cioslowski2000} - before a strong decay at long range due to the appearance of the Gaussian function. Such a behavior has been noticed before in the Wigner regime of both Hooke's atom \cite{Taut1993} and the spherium atom \cite{Loos2010b}.

\section{\label{conc}Conclusion}

We have uncovered closed form solutions for a quantum mechanical three-body system, which has a physical application as a three-dimensional helium-like atom. We have derived some exact expressions for the two-electron ground state wavefunction and its associated electronic density. We hope that this atomic model is useful for future work on fundamental issues such as entanglement and Fisher-Shannon information, as well as being a theoretical laboratory to test concepts and approximations in density functional theory and other numerical methodologies \cite{Capelle2013}. Finally, we note that the artificial atom described in this work also leads to analytical solutions in lower dimensions.

\section*{Acknowledgments}
We acknowledge financial support from the Centre national de la recherche scientifique (CNRS). We would like to thank E.Leclerc for refreshing and fruitful assistance.

\end{document}